\documentstyle[twocolumn,psfig,pre,aps]{revtex}
\begin{document}
\title{Localization-delocalization transition of a reaction-diffusion 
front near a semipermeable wall.}
\author{Bastien Chopard}
\address{Computer Science Department, Universit\'e de Gen\`eve,
CH 1211 Gen\`eve 4, Switzerland.}
\author{Michel Droz and J\'er\^ome Magnin}
\address{D\'epartement de Physique Th\'eorique, Universit\'e de Gen\`eve,
CH 1211 Gen\`eve 4, Switzerland.}
\author{Zolt\'an R\'acz}
\address{Institute for Theoretical Physics, E\"otv\"os
University, 1088 Budapest, Puskin u. 5-7, Hungary}

\date{\today}
\maketitle
\begin{abstract}
The $A+B \to C$ reaction-diffusion process is studied in a system where
the reagents are separated by a semipermeable wall.  We use
reaction-diffusion equations to describe the process and to  derive a scaling description for the
long-time behavior of the reaction front. Furthermore, we show that
a critical localization-delocalization transition takes place as a
control parameter which depends on the initial densities and on the
diffusion constants is varied.  The transition is between a reaction
front of finite width that is localized at the wall and a front which
is detached and moves away from the wall. At the critical point, the
reaction front remains at the wall but its width diverges with time (as
$t^{1/6}$ in mean-field approximation).

\vspace {0.3truecm} 
\end{abstract}
\section{Introduction.}
Reaction fronts formed in diffusion-limited $A+B \to C$ type reactions 
have been investigated intensively in recents 
years~\cite{GR,koo,Ebner,Koo,Chopard,Cornell,Taitelbaum,Leyvraz,Ben-Naim,Araujo,Larralde,Havlin,droz,cardi-uno,cardi-due,koza-uno,koza-due,koza-tre}. 
The motivation comes partly from the realization that 
moving reaction fronts play an important role in a great variety of physical 
and chemical phenomena which display pattern 
formation~\cite{langer,dee,luthi-uno,cross}. Another reason for the 
interest is the simplicity of the problem which allows the application 
of different theoretical approaches.
Indeed, the front properties have been 
studied in detail by using mean-field and scaling theories~\cite{Cornell}, 
dynamical renormalization 
group~\cite{cardi-uno}, numerical simulations~\cite{bastien}) and 
in some cases exact analytical predictions have also been made~\cite{droz}.  

In most of the cases studied previously, the reaction front is formed
after the spatially separated components $A$ and $B$ come into
contact.  For example, in a typical experiment aimed at producing
Liesegang bands~\cite{Zrinyi}, one has a vertical tube of gel soaked
with component $B$, and, at time $t=0$, a liquid containing the reagent
$A$ is poured over the gel (in order to eliminate convection effects,
the liquid is sometimes replaced with another gel containing $A$).  The
theoretical equivalent of this situation is that the reagents are
separated by a wall which is removed at $t=0$ and then the
reaction-diffusion process begins.

One can imagine, however, that there are situations when 
the wall between the reagents is present at all times, and this 
wall is semipermeable allowing only one 
of the reagents to pass through. It may happen, for example, in the above discussed 
setup that $B$ is not soluble in the liquid containing $A$ which 
is effectively equivalent to the presence of a semipermeable wall. 
More importantly, chemical reactions in biological systems take usually place 
in strongly inhomogeneous media with semipermeable walls 
present~\cite{Thecell,radiolaria,diatom}.
Thus, we believe it is important (hence the aim of this
paper) to consider the formation of reaction fronts
in systems with initial separated species when the wall separating the 
two species is not eliminated at $t=0$ but is replaced by a 
semipermeable wall which allows only one of the 
reagents ($A$) to diffuse across.

Using a mean-field description of the above process, we find 
that the control parameter in this system is given by 
\begin{equation}
r=1-\frac{b_0 \sqrt D_b}{a_0 \sqrt D_a} \label{control}
\end{equation}
where $a_0$ and $b_0$ are the initial particles densities while  
$D_a$ and $D_b$ are the  diffusion coefficients. We show that, depending 
on the sign of $r$, three distinct types of behavior occur.
When $r>0$, the $A$ particles invade the $B$ phase.
The reaction front moves away from the semipermeable wall with the distance 
from the wall increasing as $\sqrt{t}$ and
the wall is irrelevant in the long time regime.
Thus, one recovers the predictions (e.g. the width of the reaction zone 
scales as $w\sim t^{1/6}$) made with no semipermeable wall present~\cite{GR}. 
In the opposite case,
$r<0$, the wall prevents the $B$ particles from invading the $A$ region and,  
accordingly, the reaction front becomes localized (with finite width) 
at the semipermeable wall.  
It turns out that the dividing point between the $r>0$ and $r<0$ cases 
is a critical point
in the sense that the width of the reaction zone  diverges at $r=0$. 
We have thus found a critical localization-delocalization transition from
a reaction front localized at the wall to a front detached and 
moving away from the wall. 

The above results will be derived and discussed first by defining the model 
(dynamical equations and the boundary conditions) in Sec. II. Then, 
the different 
regimes are analyzed (Sec. III) both analytically and numerically 
at the mean-field level with comments on the role of the fluctuations. 
Concluding remarks are given in Sec. IV.

\section{The model} 
 \label{model}
The basic notions about reaction zones have been introduced for 
the $A+B\rightarrow C$ process \cite{GR} and, in order to keep the 
discussion transparent, we shall also consider this case. More 
complicated reaction schemes $\nu_AA+\nu_BB\rightarrow C$ can be treated along 
the same line with the same general picture arising.

We shall assume that the transport kinetics of the 
reagents is dominated by diffusion and that the reaction kinetics 
is of second order. Thus, at a mean-field level, 
the mathematical description of the process is given in terms of 
reaction-diffusion equations
\begin{eqnarray}
\partial_t a=&&D_a\nabla^2 a-kab \label{A2} \quad ,\\
\partial_t b=&&D_b\nabla^2 b-kab \label{B2} \quad ,
\end{eqnarray}
where $a$ and $b$ are the densities of the reagents $A$ and
$B$, respectively, $D_a$ and $D_b$ are the corresponding diffusion constants,
and the reaction-rate parameter is $k$. Note that there is a conservation law 
in this system. Since the $A$ 
and $B$ particles react in pairs the difference in their numbers is  
conserved. In terms of the densities this means that the 
spatial integral of $a-b$ is constant unless there are particle sources 
at the boundaries. 

The semipermeable membrane is located at the $(x=0,y,z)$ plane. 
Initially, all $B$ particles are 
on the right hand side of this membrane ($x>0$) and, since 
the membrane is impenetrable for 
them, they remain on that side for all times. In terms of the 
particle density $b$ this means that the solution of (\ref{A2}) and (\ref{B2})
must satisfy the following conditions
\begin{equation}
b(x<0,t)=0\quad , \quad 
\left.\frac{\partial b(x,t)}{\partial x}\right |_{x=0^+}=0 
\quad .
\label{boundatx=0}
\end{equation}

The motion of the $A$ particles is not influenced by the membrane and, 
initially, they are on the left side of it. 
Furthermore, the initial densities are assumed to be constant i.e.
$a(x,0)=a_0$ and $b(x,0)=0$ for $x<0$ while 
$a(x,0)=0$ and $b(x,0)=b_0$ for $x>0$.
With this choice of initial state, the solution of (\ref{A2}) and (\ref{B2}) 
depends only on the $x$ spatial coordinate and the system effectively
becomes one-dimensional.

Our aim will be to calculate the production rate of $C$ particles defined by 
\begin{equation}
R(x,t)=ka(x,t)b(x,t) \quad ,
\label{R(x,t)}
\end{equation}
and investigate the time-evolution of its spatial structure with emphasis 
on the center 
\begin{equation}
x_f(t)={\int_{-\infty}^{\infty}xR(x,t)dx}/
{\int_{-\infty}^{\infty}R(x,t)dx}\quad 
\label{center}
\end{equation}
and the width of the reaction zone 
\begin{equation}
w(t)=\left[{\int_{-\infty}^{\infty}(x-x_f)^2R(x,t)dx}/
{\int_{-\infty}^{\infty}R(x,t)dx}\right]^{1/2}.
\label{width}
\end{equation}
Both $x_f$ and $w$ are the easily measurable quantities in 
experiments and simulations.
	
\section{Scaling properties of the front} 
 \label{Scaling}
 
For a system without the membrane, it is known~\cite{Ebner,koza-due} that 
the reaction front will move to the right ($A$ invades $B$) or to the 
left ($B$ invades $A$) depending on the relative magnitude of 
quasistationary diffusive currents
($J^A\sim D_aa_0/\sqrt{D_at}$ and $J^B\sim D_bb_0/\sqrt{D_bt}$),
i.e. depending on the sign of the 
control parameter $r$:\begin{equation}
r=1-\frac{J^B}{J^A}=1-\frac{b_0\sqrt{D_b}}{a_0\sqrt{D_a}}\quad .
\label{diszkrim}
\end{equation}
For $r=r_c=0$, the front is stationary in the sense that although $R(x,t)$ 
remains time-dependent for large times, 
the center of the reaction zone does not move and $x_f(t\rightarrow \infty)$ 
approaches a finite constant.

One expects that the direction of invasion plays an important role 
in the presence of the membrane as well and, accordingly, we shall analyze 
the $r>0$, $r<0$, and $r=0$ cases separately.

\subsection{$r>0$: Invasion of the {\it free} ($A$) reagents -- 
delocalized front}
 \label{free}

\begin{figure}
\centerline{\psfig{file=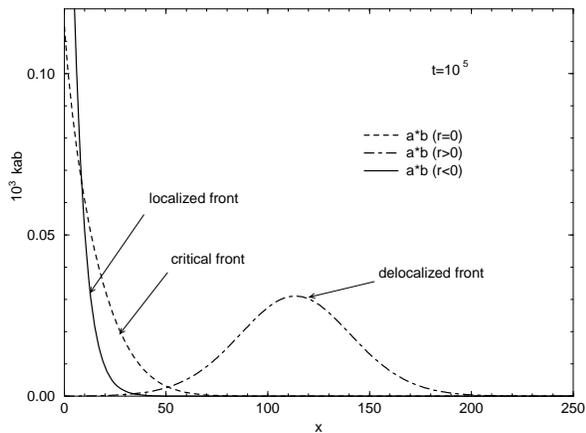,width=8cm}}
\caption{Production rate of $C$ near the semipermeable membrane 
(localized at $x=0$) for three different values of the control parameter
$r$.  The width of the localized front is stationary while the width of the
critical front increases with time as $t^{1/6}$. The distance between the
delocalized reaction front and the membrane increases as $t^{1/2}$ while
its width diverges as $t^{1/6}$.}
\label{fig1}
\end{figure}

For $r>0$, the diffusive current of $A$ particles overwhelms the corresponding
current of $B$ particles and thus the reaction front moves to the right.
After a while, the $B$ particles disappear from the 
neighborhood of the membrane and thus the membrane does not play a role 
anymore. Consequently, the reaction front leaves the membrane (Fig.\ref{fig1})
and all the results about the long-time 
scaling form of the reaction front obtained previously apply~\cite{GR}, 
namely
\begin{equation}
R(x,t)\sim t^{-\beta}F\left({x-x_f}\over t^\alpha \right)
\quad ,
\label{R(x,t)scale}
\end{equation}
where the position of the center of the  front, $x_f$,  scales  with 
time as $x_f\sim \sqrt{t}$, 
the width of the reaction front is proportional  to $t^\alpha$  with 
$\alpha=1/6$ , the scaling exponent of the production rate of $C$ at 
$x=x_f$ is $\beta=2/3$, and the scaling function, $F(z)$, is a fast 
decreasing function for $z\rightarrow\pm\infty$.

We can call this front delocalized since both the center and the 
width of the front diverge in the long-time limit.

In closing this subsection, we note that 
the above results are modified by fluctuations in low dimensions ($d<2$), 
as discussed in several works on $A+B\rightarrow C$ reactions 
without the presence of a membrane 
\cite{Cornell,droz}. 

\subsection{$r<0$: Invasion of the {\it blocked} ($B$) reagents -- 
localized front}
 \label{blocked}
\begin{figure}
\centerline{\psfig{file=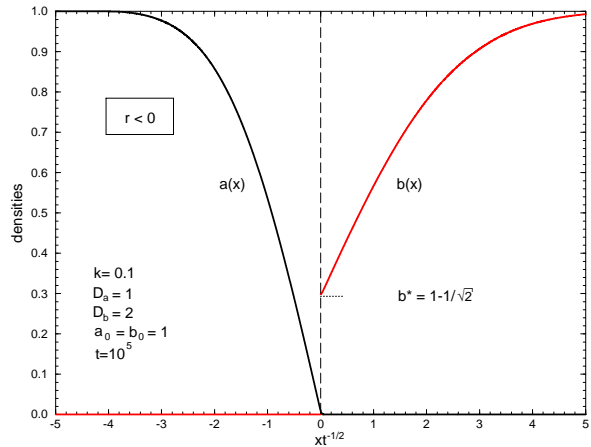,width=8cm}}
\caption{ Density profile of the reagents for $r<0$ as seen on a 
diffusive scale ($x \sim t^{1/2}$). Time is measured in units of $\tau=0.1/(ka_0)$ where $k$ is the reaction rate and $a_0$ is the initial density of $A$. The unit of length is chosen to be $\ell=\sqrt{D_a \tau}$ where $D_a$ is the diffusion coefficient of $A$. For the given values of diffusion coefficients ($D_a,D_b$) and initial densities ($a_0, b_0$), the large time limit of $b$ at $x=0$ is given by $b^\star=1-1/\sqrt{2}$. }
\label{fig2}
\end{figure}

For $r<0$, the $B$ particles would be the invading particles but they 
cannot penetrate past the membrane. Thus, one expects that there will 
be a finite density of $B$ particles at $x=0$ and, consequently, the 
$A$ particles can penetrate into 
the $x>0$ region only up to a finite distance, $\xi$. In order to make this 
picture (Fig.\ref{fig2}) quantitative, we shall first solve the problem on the diffusive 
lengthscale $x\sim \sqrt{t}$ and then use this solution 
as the large-argument asymptotics of the solution around $x=0$.

Viewing the process on the diffusive lengthscale, the reaction zone is 
reduced to a point ($x=0$) and the diffusion of $A$ and $B$ takes place 
separately in the $x<0$ and $x>0$ regions. The appropriate 
boundary conditions are as follows:
\begin{equation}
a(x\rightarrow -\infty,t)=a_0\quad , \quad a(0,t)=0\quad , 
\label{boundcond1}
\end{equation}
\begin{equation}
b(x\rightarrow \infty,t)=b_0\quad , 
\quad -D_a\left. \frac{\partial a}{\partial x}\right|_{x=0^-}=
D_b\left. \frac{\partial b}{\partial x}\right|_{x=0^+}
.
\label{boundcond2}
\end{equation}
The first boundary conditions in (\ref{boundcond1}) and (\ref{boundcond2}) are
obvious while the second boundary condition in (\ref{boundcond2}) 
is just the expression of the
equality of the currents entering the reaction zone.
The second condition in (\ref{boundcond1}) is more complicated. 
It follows from the assumption that the penetration length, $\xi$, 
is finite combined with the fact that 
the diffusion current approaches zero at large times,
[$J_{diff}\sim D_a(\partial a/\partial x)|_{x=0}\rightarrow 1/\sqrt{t}$ i.e.
the derivative $(\partial a/\partial x)|_{x=0}$ 
diminishes for $t\rightarrow \infty$].
The finiteness of $\xi$,
in turn, follows from the finiteness of $b(0,t)=b^*$ and so, 
finding $b^*$ finite at the end of our calculation provides a 
selfconsistency check of the underlying picture. 

The solution of the diffusion equations with the above boundary conditions 
is given by 
\begin{eqnarray}
a(x,t)=&&-a_0{\rm Erf}(x/\sqrt{4D_at}) \label{solution_a} \quad ,\\
b(x,t)=&&b^*+(b_0-b^*){\rm Erf}(x/\sqrt{4D_bt}) \label{solution_b}
\quad ,
\end{eqnarray}
where Erf$(x)$ is the error function
~\cite{abramo} and 
$b^*=b(0,t)$ is found from the second condition 
in (\ref{boundcond2}):
\begin{equation}
b^*=b_0-a_0\sqrt{\frac{D_a}{D_b}}=-a_0\sqrt{\frac{D_a}{D_b}}r
\quad .
\label{b*}
\end{equation}
As we can see, $b^*$ is indeed finite for finite $r<0$ ($b^*>0$ because it 
has the meaning of particle density).

\begin{figure}
\centerline{\psfig{file=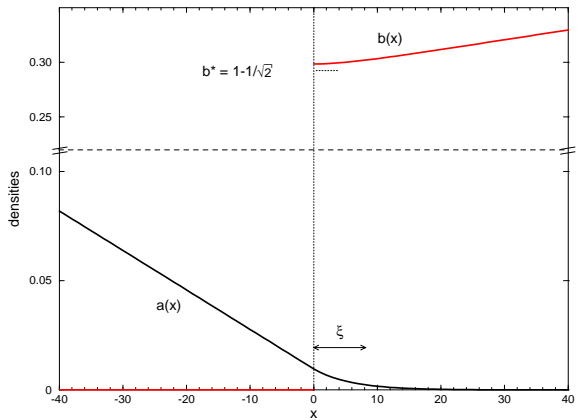,width=8cm}}
\caption{ Magnified view of the reaction zone shown in Fig.2. Here
the $x$ coordinate is not scaled by $t^{1/2}$. The penetration length of 
particles $A$ into the $B$ region is shown by $\xi$. 
Note that there is a break in the vertical scale.}
\label{fig3}
\end{figure}

The above results are valid on lengthscale $x\sim t^{1/2}$. In order to 
investigate the details of the reaction zone, 
we must consider the $x\sim t^0$ region (Fig.\ref{fig3})
where we should find a solution with large-distance asymptotics 
which matches to the solution found above. 

Since we are mainly interested to find the extent of the region
where the reaction product appears, we should find the region of 
penetration of $A$ particles into the $x>0$ halfspace. For $x\ll \sqrt{t}$, 
one can approximate $b(x,t)\approx b^*$ and then the equation for $a$ becomes linear 
\begin{equation}
\partial_t a=D_a\nabla^2 a-kb^*a  \quad .
\label{Alinear}
\end{equation}
This equation is supplemented with the following boundary conditions
\begin{equation}
a(x\rightarrow\infty,t)=0 \quad , \quad 
\left.\frac{\partial a}{\partial x}\right |_{x=0}=-\frac{a_0}{\sqrt{\pi D_at}}  
\quad .
\label{Aboundary}
\end{equation}
The second condition comes from the fact that 
the diffusion current entering the reaction zone
at $x=0$ must be equal to that calculated from the macroscopic $(x\sim\sqrt{t})$
considerations.

Due to the slowness of diffusion, $a(x,t)$ changes slowly at large times 
and one can consider quasistatic approximation. Looking for a solution of the 
form
\begin{equation}
a(x,t)\approx\frac{1}{\sqrt{t}}\Phi(x)
\quad ,
\label{Fi}
\end{equation}
one can see that the left hand side of (\ref{Alinear}) is of the order 
$t^{-3/2}$ while the right hand side is proportional to $t^{-1/2}$ and so, the 
time derivative can be neglected.
The resulting equation for $\Phi$ can be easily solved and the boundary 
conditions can be satisfied yielding a solution in a scaling form:
\begin{equation}
\frac{a(x,t)}{a_0}=\Psi(x/\xi, D_at/\xi^2)=
\frac{e^{-\frac{x}{\xi}}}{\sqrt{\pi D_at/\xi^2}}
\quad ,
\label{scalingA}
\end{equation}
where the penetration (or correlation) length is given by
\begin{equation}
\xi=\sqrt{\frac{D_a}{kb^*}}\sim|r|^{-1/2}
\quad .
\label{penetlength}
\end{equation}
Since $b(x,t)\approx b^*$ in the reaction zone, we can obtain $R(x,t)$ 
from (\ref{scalingA}):
\begin{eqnarray}
R(x,t)=kab&& \approx kab^*\sim \frac{a_0D_a}{\sqrt{\pi D_at}}
\frac{e^{-\frac{x}{\xi}}}{\xi} \quad x>0\quad ,\\
&&=0\quad \quad\quad \quad \quad\quad \quad \quad \quad \, x<0\quad .
\label{Rscaling}
\end{eqnarray}

\begin{figure}
\centerline{\psfig{file=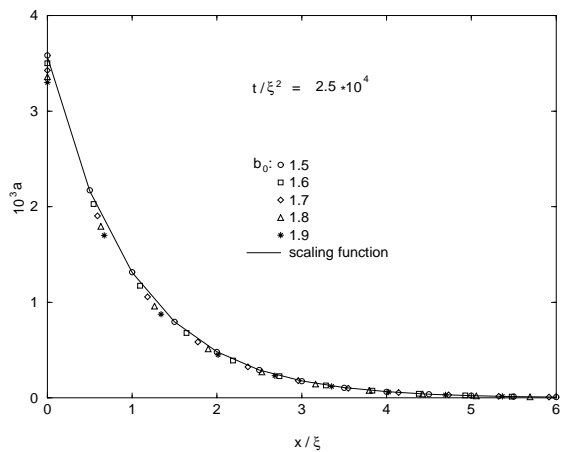,width=8cm}}
\caption{ Scaling of the density of $A$'s in the reaction zone shown in Fig.~2.
The parameters are the same 
as in Fig.~2 except for $t$ and $b_0$ which are varied in order to
keep $t/\xi^2$ constant [$\xi$ is given by equation (19)].
The numerical solution of the full set of reaction-diffusion equations  (equations (2) and (3)) is compared with the 
quasi-stationary scaling solution, $\Phi_a$ (solid line). 
Since one has $b \approx b^*$ in the reaction zone, the scaling function 
of the reaction rate $R=kab= \approx kab^*$ is proportional to  
that of $a$.}
\label{fig4}
\end{figure}

Thus the reaction rate goes down with time as $1/\sqrt{t}$ while
the center and the width of the reaction zone remain finite in this scaling 
limit
\begin{equation}
x_f\sim w\sim \xi\quad .
\label{xwscaling}
\end{equation}
One can see from Fig.\ref{fig4} that the scaling form (\ref{scalingA}) agrees
with the scaling obtained from the numerical solution of the full set of 
equations (\ref{A2}) and (\ref{B2}).

The phase considered above may be
called the phase of localized reaction zone. One can observe from 
(\ref{penetlength}), however, that $\xi$ diverges as we approach 
the $r=0$ point and thus the reaction zone becomes delocalized at 
$r=r_c=0$. Thus 
$r$ plays the role of the distance from a critical point and the 
exponent we found, $\xi\sim r^{-\nu}\sim r^{-1/2}$ is obviously the 
mean-field exponent $\nu=1/2$ in accord with the neglect of fluctuations
in the above description.

\subsection{$r=0$: Localization-delocalization transition -- critical front}
 \label{critical}

\begin{figure}
\centerline{\psfig{file=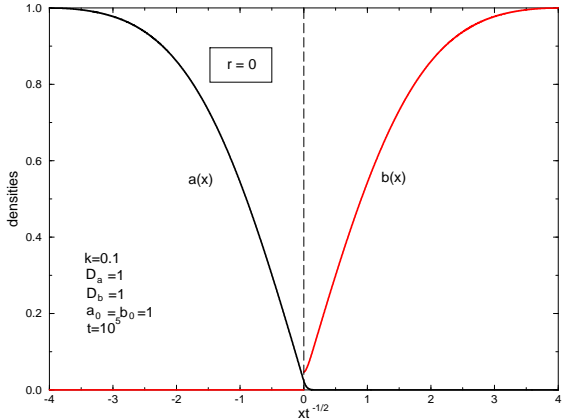,width=8cm}}
\caption{Density profile of the reagents at the critical point ($r=0$) as seen on a diffusive  scale ($x \sim t^{1/2}$). Notation is explained in caption  to Fig. 2.}
\label{fig5}
\end{figure}

\begin{figure}
\centerline{\psfig{file=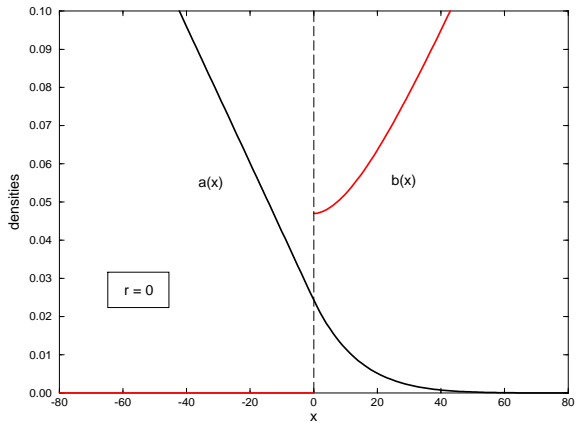,width=8cm}}
\caption{Magnified view of the reaction zone shown in Fig.5. Note that here $x$ 
is not scaled by $t^{1/2}$.}
\label{fig6}
\end{figure}

It follows from the previous subsection that the $r=0$ case can be considered 
as a critical point which separates the localized and delocalized phases. 
Thus we expect that a scaling description is valid 
again at $r=r_c=0$ but, in expressions like (\ref{scalingA}), 
the correlation length must be replaced by a time-dependent correlation length
which scales as a power of time, $\xi(t)\sim t^\alpha$. In order to see that 
this picture is valid, we follow the steps of the previous subsection: 
the problem is first solved on the diffusion scale [the solution is actually given by 
equations (\ref{solution_a}) and (\ref{solution_b})  with $b^*=0$] and then 
matching solution in the $x\approx 0$ region is found 
(Fig.\ref{fig5} and \ref{fig6}).

In the $x\approx 0$ region we seek scaling solutions suggested by 
eq.(\ref{scalingA})
\begin{equation}
a(x,t)\approx
\frac{\Phi_a(x/t^\alpha)}{t^{1/2-\alpha}}\quad , \quad 
b(x,t)\approx
\frac{\Phi_b(x/t^\alpha)}{t^{1/2-\alpha}}
\label{scalingAB}
\end{equation}
Several comments are in order to clarify the above scaling assumptions.
First, the scaling of $x$ by the same $t^\alpha$ in $\Phi_a$ and $\Phi_b$ 
is the assumption that there 
is only one lengthscale governing the reaction zone. Second, 
the exponent $\alpha$ should be $1/6$ or less since 
the case without the membrane gives an upper limit for the spread of the 
reaction zone and there the width is proportional to $t^{1/6}$. 
Finally, one should 
note that the exponent, $1/2-\alpha$, of the prefactors of the 
scaling functions is, in principle, an independent exponent. In this 
case, however, it is fixed by the boundary condition 
$(\partial a/\partial x)_{x=0}\sim 1/\sqrt{t}$ and by the requirement that 
the large argument asymptotics of $b(x,t)$ should match the solution 
obtained on the $x\sim\sqrt{t}$ scale. 

Substituting the scaling forms (\ref{scalingAB}) into equations 
(\ref{A2},\ref{B2}), one finds that, for large times and for $\alpha<1/2$, 
the time derivatives on the
left hand sides can be neglected. Furthermore, 
the right-hand sides yield meaningful equations only if 
$\alpha$ is set to $\alpha=1/6$. The resulting equations then take the form
\begin{eqnarray}
\frac{d^2\Phi_a}{dz^2}=&&\frac{k}{D_a}\Phi_a\Phi_b \label{PhiA} \\
\frac{d^2\Phi_b}{dz^2}=&&\frac{k}{D_b}\Phi_a\Phi_b \label{PhiB} \quad ,
\end{eqnarray}
where the scaling variable is $z=x/t^{1/6}$.

The boundary conditions to the above equations follow from 
$a(x\rightarrow\infty,t)=0$ and $\partial b/\partial x(0,t)=0$ and from 
matching the solutions to the ones found on the diffusive scale
\begin{eqnarray}
\Phi_a(z\rightarrow\infty)=0\quad &&, \quad 
\left.\frac{d\Phi_a}{dz}\right |_{z=0}=-\frac{a_0}{\sqrt{\pi D_a}} 
\label{PhiboundA} \quad ,\\
\left.\frac{d\Phi_b}{dz}\right |_{z=0}=0\quad &&, \quad
\left.\frac{d\Phi_b}{dz}\right|_{z\rightarrow \infty}=\frac{b_0}{\sqrt{\pi D_b}} 
\label{PhiboundB}\quad .
\end{eqnarray}
Equations (\ref{PhiA},\ref{PhiB}) with boundary conditions 
(\ref{PhiboundA},\ref{PhiboundB}), however, pose a difficulty 
related to the fact that the combination
$v=D_a\Phi_a-D_b\Phi_b$
satisfies a linear equation $v^{\prime\prime}=0$ and the solution, $v=Pz+Q$, 
contains a integration constant, $Q$, which is not determined 
by the boundary conditions. Consequently, the scaling functions 
do not appear to be unique.

This problem of uniqueness can be dealt with by returning to the 
diffusive scale, $x\sim\sqrt{t}$, and reexamining the solutions found there. 
We shall demonstrate the idea on the example of a system where $D_a=D_b$
(and $a_0=b_0$ since we are at criticality). In this case, $u=a-b$ 
satisfies the diffusion equation for both $x>0$ and $x<0$, 
the boundary conditions are given 
by $u(-\infty,t)=-u(\infty,t)=a_0$ and 
$\partial_xu(0^-,t)=\partial_xu(0^+,t)$ and, furthermore, the initial 
condition [$u(x<0,0)=a_0$; $u(x>0,0)=-a_0$] is an odd function of $x$.
It follows then that the solution is an odd function,
$u(x,t)=-u(-x,t)$. Next we note that $u=a$ for $x<0$ while $u=a-b$ for $x>0$ 
and approaching $x=0$ from both sides, the oddness of $u$ yields the 
following relationship 
\begin{equation}
-a(x=0^-,t)=a(x=0^+,t)-b(x=0^+,t) \quad .
\label{boundvalues}
\end{equation}
Since there is no accumulation of $A$ particles at $x=0$, we have 
$\partial a(x=0^-,t)= \partial a(x=0^+,t)$ and, consequently, 
$a$ is continuous function across the membrane, $a(x=0^-,t)=a(x=0^+,t)$. 
Then equation (\ref{boundvalues}) yields $b(0,t)=2a(0,t)$ which, in turn, 
provides an additional boundary condition for the scaling functions:
\begin{equation}
\Phi_b(0)=2\Phi_a(0) \quad .
\label{scboundvalues}
\end{equation}
The same extra boundary condition can also be found for 
$D_a\not= D_b$ but the argument is 
rather involved, so
we shall not reproduce it here. From the perspective of
critical phenomena, it is quite natural that the scaling function does not
depend on such details as the diffusion coefficients.

Having the extra boundary condition (\ref{scboundvalues}), 
$\Phi_a$ and $\Phi_b$ can now be found numerically. 
Some properties of the scaling functions can, however, be seen by 
just inspecting the equations. For example, substituting the large $z$ 
asymptotics $\Phi_b(z)\sim z$ into (\ref{PhiA}), one can see that 
$\Phi_a(z\rightarrow\infty)$ is given by the Airy function \cite{abramo}.

\begin{figure}
\centerline{\psfig{file=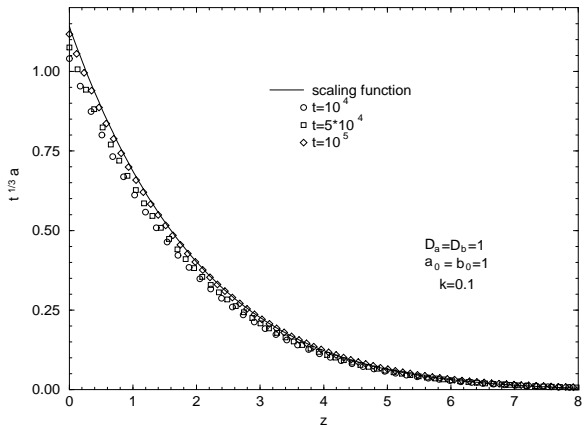,width=8cm}}
\caption{Scaling function, $\Phi_a$  for $a$ (equation (23)). The numerical solution of  the full set of reaction-diffusion equations (equations (2) and (3)) is compared with the quasi-stationary scaling solution.}  
\label{fig7}
\end{figure}

On Fig.\ref{fig7}, we show that the scaling regime does exist and that 
the numerical results
do agree with the solution of the full equations (\ref{A2},\ref{B2}).

It follows then from (\ref{PhiA},\ref{PhiB}) that the production rate can 
also be written in a scaling form
\begin{equation}
R(x,t)\sim \frac{1}{t^{2/3}}\Phi_a\left(\frac{x}{t^{1/6}}\right)
\Phi_b\left(\frac{x}{t^{1/6}}\right) 
=\frac{1}{t^{2/3}}\Psi\left(\frac{x}{t^{1/6}}\right) 
\label{scAB}
\end{equation}
and we can observe that the reaction front 
remains attached to the wall but 
it expands with time into the $x>0$ region. 
Both the center and the width of the zone diverge with time as
\begin{equation}
x_f\sim w\sim t^{1/6}\quad .
\label{xwsc}
\end{equation}
and both exponents are the same in contrast to the delocalized phase where 
$x_f\sim t^{1/2}$ and $w\sim t^{1/6}$.

The above discussion is based on a mean-field like treatment and 
one can ask what is the role of the fluctuations which are supposed to 
be important in low dimensions.
Numerical simulations carried out
for an one- and two-dimensional systems with semipermeable wall ~\cite{tpa}
indicate that the mean field description discussed above 
remains qualitatively correct. The upper critical dimension above 
which the mean field theory is correct appears to be $d_u=2$. 
However, in dimension $d=1$, the critical exponents takes their non mean 
field values~\cite{droz}. For example, the mean position and 
width exponents at the critical point change from the mean field value 
$\alpha=1/6$ to $\alpha=1/4$.

We can now summarize the properties of the loca\-lization-delocalization
transition discussed above as follows. 
For $r<0$, the reaction zone is localized 
at the membrane and the width is determined 
by the correlation length, $\xi$, describing the penetration 
of the A particles into the constant-density $B$ region. 
At $r=0$ the penetration length diverges but
there is still a single (diverging with time) length which characterizes 
the reaction zone. It should be noted
that a diverging diffusion length $\ell_D\sim \sqrt{t}$ is always present 
but it is irrelevant for $r\le0$. 
For $r>0$, however, the diffusion length starts to play a role: 
the reaction zone gets delocalized and two distinct lengthscales appear. 
One of them is the distance of 
the center of the zone from the membrane, $x_f\sim\sqrt{t}$, which is just the 
diffusion length while the other
is the width of the reaction zone, $w\sim t^{1/6}$.

\section{Final remarks} 
 \label{Final}

The questions of how much $C$ is produced near the membrane 
and whether their density, $c$, grows to exceed
some aggregation threshold, $c_0$, may be of importance in
biological phenomena (e.g. in the building of rather intricate 
but regular mineral skeletons of single-cell organisms such as 
radiolaria~\cite{radiolaria} or diatoms~\cite{diatom}). 
The answers to the above questions depend on the 
localization properties of the reaction zone. 

For $r<0$, the reaction zone has a finite width and thus, 
provided the $C$-s do not diffuse away,
their density will increase with time as $c(t)\sim \sqrt{t}$. 
This result follows from 
the fact that the current, $J^A(t)$, of $A$ particles towards the reaction 
zone is proportional to $1/\sqrt{t}$ and, consequently, 
the amount of $C$-s, produced 
up to time $t$, is given by $M_C\sim\int^tJ^A(\tau)d\tau\sim \sqrt{t}$. 

A somewhat slower increase of $c(t)$ takes place at $r=0$. Since 
the width of the reaction zone diverges as $w\sim t^{1/6}$,
one finds $c(t)\sim M_C/w\sim t^{1/3}$. We can see that, for both $r< 0$ 
and $r=0$, the density of $C$-s near the membrane exceeds 
any threshold $c_0$ at sufficiently large times. Thus 
supersaturation and, associated with it, the precipitation 
of $C$ may occur in these regimes.

Finally, for $r>0$, 
the reaction zone leaves the membrane and only a finite density of 
$C$-s left behind. The actual value of this density depends sensitively 
on the initial conditions and one cannot make statements about 
possible precipitation without knowledge of the actual parameters.

The above considerations, of course, do not constitute an attempt towards 
the explanation of a real biological phenomena such as the 
precipitation of the siliceous stuctures of single-cell 
radiolaria. This is so even if one imagines that, 
at the early stages of the evolution, the regular 
skeletons are either produced as an instability in a physico-chemical, 
reaction-diffusion process or arose by  
surface-tension assisted precipitation where the membranes 
are present but play a passive role (their intersections defines the 
precipitation regions)~\cite{Darcy}. 
At present stage of evolution,  
the skeletons are covered with a membranous cytoplasmic sheat which appears to 
play an important role (e.g. transport along the 
membrane) in the skeletal depositions\cite{radiolaria}. Thus any 
attempt at {\em physico-chemical} explanation should include the presence 
of such an {\em active} membrane near the precipitation zone. 

In this paper, we have derived results for the properties of 
reaction zones near a 
semipermeable membrane which is {\em active} only in the sense
that it is blocking the transport of one of the reagents. 
We hope, however, that our results will help 
discussing more complicated reactions near {\em active} membranes 
in the same way as the 
understanding of the properties of the reaction zone~\cite{GR} in the 
$A+B\rightarrow C$ reaction helped 
in elucidating the features of the pattern formation 
in the much more complicated Liesegang phenomena~\cite{luthi-uno}.

\section*{Acknowledgments}
This work has been partially supported by the Swiss National Science Foundation 
in the framework of the Cooperation in Science and Research with CEEC/NIS, 
by the Hungarian Academy of Sciences (Grant OTKA T 019451),
and by an EC Network Grant ERB CHRX-CT92-0063. Z.R. would like to thank 
for the hospitality of the members of the Theoretical Physics Department 
during his stay at the University of Geneva.






\begin{references}
\bibitem{GR} L. G\'alfi and Z. R\'acz, Phys. Rev. {\bf A38}, 3151 (1988).
\bibitem{koo} Y.E. Lee Koo, L. Li, and R. Kopelman, 
Mol. Cryst. Liq. Cryst. {\bf 183}, 187 (1990).
\bibitem{Ebner} Z. Jiang and C. Ebner, Phys. Rev.{\bf A42}, 7483 (1990).
\bibitem{Koo} Y-E. Lee Koo and R. Kopelman, J. Stat. Phys. {\bf 65}, 893 (1991).
\bibitem{Chopard} B. Chopard and M. Droz, Europhys. Lett. {\bf 15}, 459 (1991).
\bibitem{Cornell} S. Cornell, B. Chopard and M. Droz, Phys. Rev. {\bf A44} 4826, (1991).
\bibitem{Taitelbaum} H. Taitelbaum, S. Havlin, J. E. Kiefer, B. Trus, and 
G. H. Weiss, J. Stat. Phys. {\bf 65}, 873 (1991).
\bibitem{Leyvraz} F. Leyvraz and S. Redner, Phys. Rev. Lett. {\bf 66}, 2168 (1991); J.Stat.Phys. {\bf 65}, 1043 (1991).
\bibitem{Ben-Naim} E. Ben-Naim and S. Redner, J.Phys. {\bf A25}, L575 (1992).
\bibitem{Araujo} M. Araujo, S. Havlin. H. Larralde, and H.E. Stanley,
Phys. Rev. Lett. {\bf 68}, 1791 (1992).
\bibitem{Larralde} H. Larralde, M. Araujo, S. Havlin, and H.E. Stanley.
Phys. Rev. {\bf A46}, 855 (1992).
\bibitem{Havlin} H. Taitelbaum, Y.E. Lee Koo, S. Havlin, R. Kopelman,
and G.H. Weiss, Phys. Rev. {\bf A46}, 2151 (1992).
\bibitem{droz} S. Cornell and M. Droz, Phys. Rev. Lett. {\bf 70}, 3824 (1993).
\bibitem{cardi-uno} M. Howard and J. Cardy, J. Phys. {\bf A28}, 3599 (1995). 
\bibitem{cardi-due} G.T. Barkema, M.J. Howard and J.L. Cardy, 
Phys. Rev. {\bf E53}, R2017 (1996).
\bibitem{koza-uno} S. Cornell, Z. Koza  and M. Droz, 
Phys. Rev. {\bf E52}, 3500 (1995).
\bibitem{koza-due} Z. Koza and H. Taitelbaum,  
Phys.Rev. {\bf E54}, R1040 (1996).
\bibitem{koza-tre}  Z. Koza, J. Stat. Phys. {\bf 85}, 179 (1996).
\bibitem{Cabarcos} E.L. Cabarcos, C-S. Kuo, A. Scala, and R. Bansil,
Phys. Rev. Lett. {\bf 77}, 2834 (1996).
\bibitem{langer} J.S. Langer, Rev. Mod. Phys. {\bf 52}, 1 (1980).
\bibitem{dee}  G.T. Dee, J. Stat. Phys. {\bf 39}, 705 (1985); Phys.Rev.Lett.
{\bf 57}, 275 (1986).
\bibitem{luthi-uno} B. Chopard, P. Luthi, and M. Droz, 
Phys. Rev. Lett. {\bf 72}, 1384 (1994); J. Stat. Phys {\bf 76}, 661 (1994).
\bibitem{cross} M.C. Cross and P.C. Hohenberg, 
Rev. Mod. Phys. {\bf 65}, 851 (1993).
\bibitem{bastien} B. Chopard, M. Droz, and L. Frachebourg, 
Int. J. Mod. Physics, {\bf 5}, 47, (1994).
\bibitem{Zrinyi} A. B\"uki, \'E. K\'arp\'ati-Smidr\'oczki, and M. Zr\'\i nyi,
J. Chem. Phys. {\bf 103}, 10387 (1995). 
\bibitem{Thecell} D.W. Fawcett, {\em The cell} (W.B. Saunders, London, 1981).
\bibitem{radiolaria} O.R. Anderson, {\em Radiolaria} (Springer-Verlag, 
New York, 1983).
\bibitem{diatom} {\em Silicon and Siliceous Structures in Biological Systems},
Eds. T.L. Simpson and B.E. Volcani (Springer-Verlag, New York, 1981).
\bibitem{abramo}{\em Handbook of Mathematical Functions}, 
Edited by M. Abramowitz and I.A. Stegun (Dover, New York, 1965).
\bibitem{tpa} D. Roussin,  O. von Suzini, and M. Droz, (unpublished).
\bibitem{Darcy} D'Arcy W. Thompson, {\em On Growth and Form},
(Macmillan, New York, 1942).
\end{references}
\end{document}